\newcommand{\NP}{{\em Nucl.\ Phys.\ }}
\newcommand{\PL}{{\em Phys.\ Lett.\ }}

\newcommand{\PRL}{{\em Phys.\ Rev.\ Lett.\ }}

\documentstyle[12pt]{article}

\begin{document}
\begin{titlepage}
\begin{center}

\hfill       PUPT-1704\\
\hfill hep-th/9705116\\

\vskip .9in

{{\Large \bf Adhering 0-branes to 6-branes and 8-branes}}

\vskip .25in

Washington Taylor IV\\
{\small \sl Department of Physics,} \\
{\small \sl Princeton University,} \\
{\small \sl Princeton, New Jersey 08544, U.S.A.} \\
{\small \tt wati@princeton.edu} \\[0.3cm]
\end{center}
\vskip .25in

\begin{abstract}
A Yang-Mills solution is constructed on $T^6$ which
corresponds to a brane configuration composed purely of 0-branes and
6-branes.  This configuration breaks all supersymmetries and has an
energy greater than the sum of the energies of its components;
nonetheless, the configuration is stable classically, at least to
quadratic order.  An analogous construction is also given for a system
of 0-branes and 8-branes on $T^8$.  These constructions may prove
to be useful for describing 6-branes and 8-branes in M(atrix) theory.
\end{abstract}

\end{titlepage}
\newpage
\renewcommand{\thepage}{\arabic{page}}
\setcounter{page}{1}

\section{Introduction}

In the last several years, the discovery that p-brane solitons in
string theory can be simply described in terms of Dirichlet membranes
\cite{Polchinski} has led to a series of remarkable advances in our
understanding of string theory.  Among other things, D-branes have
been used to construct configurations analogous to black holes
\cite{Strominger-Vafa,Callan-Maldacena}, providing a framework in which
one can begin to use string theory to address unresolved questions
about information and entropy in black holes.

One of the most important features of D-branes which has been
exploited in the recent developments is the tendency of D-branes of
various types to bind together into supersymmetric BPS saturated
states.  Such states are generally either ``truly'' bound, in which
case the energy of the bound state is less than the sum of the
energies of the constituent D-branes, or ``marginally bound'', in
which case the energy of the bound state is equal to the sum of the
energies of the constituents.  A variety of systems of intersecting
D-branes and D-branes at angles have been considered, and their bound
states studied.  (For a review of some of these developments, see
\cite{Polchinski-TASI}.)

In this paper we consider a configuration composed purely of 0-branes
and 6-branes.  In general, a pointlike 0-brane placed on or near a
6-brane gives rise to a configuration with no supersymmetry.  The
0-brane and 6-brane tend to repel one another.  It might be expected,
therefore, that one could not find a stable configuration comprised
only of 0-branes and 6-branes.  Nonetheless, we find that by
describing a set of four 0-branes which have been smeared out over
four 6-branes wrapped on $T^6$ in terms of a gauge field with constant
curvature, it is possible to construct a configuration with only
0-branes and 6-branes which satisfies the classical equations of
motion and which is classically stable to quadratic order.  This
construction is related through T-duality to a system of four 3-branes
wrapped diagonally around a torus in such a way that each pair of
3-branes is coincident along a single direction.  Such a configuration
is similar to a system of 3-branes used in
\cite{Vijay-Finn} to study black holes in four dimensions.  It was
recently shown that a black hole supergravity solution exists with
only 0-brane and 6-brane charges \cite{Khuri-Ortin,Sheinblatt}.  Presumably at
large distances the brane configuration described here would look
precisely like such a black hole.

In Section 2, we review constructions of 0-branes bound to 2-branes
and 4-branes.  In Section 3 we describe the construction of the system
of 0-branes and 6-branes.  This construction is carried out using
methods completely analogous to those used in the lower dimensional
examples in Section 2.  In Section 4 we give a brief description of an
analogous construction for a system of 0-branes and 8-branes.

A note on conventions: throughout this paper we normalize energies so
that in any given configuration the brane of largest dimension has
energy equal to its volume.  Thus, for example, after compactification
on a $T^2$ with unit volume the energy of a 2-brane is 1 and the
energy of a 0-brane is $4 \pi^2 \alpha'$.

\section{0-branes on 2-branes and 4-branes}

In this section we briefly review the gauge theory descriptions of
bound states of 0-branes with 2-branes and 4-branes on tori.  For more
details see, for example, \cite{Polchinski-TASI,Sanjaye-Zack,ht}.  
(Note that the binding of 0-branes with 0-branes is significantly more
subtle \cite{Sethi-Stern}.)
A
system of $N$ coincident Dirichlet $(2 p)$-branes is described in the
low-energy regime by a $U(N)$ super Yang-Mills theory in $2 p + 1$
dimensions, which can be described as the dimensional reduction of
$N=1$ 10D SYM \cite{Witten-bound}.  In the Yang-Mills theory, a
0-brane attached to the $(2 p)$-branes is described by a configuration
of the gauge fields with a unit of topological charge proportional to
the integral of $F^{\wedge p} = F \wedge \cdots \wedge F$
\cite{Witten-small,Douglas}.

\subsection{0-branes on 2-branes}

A $U(N)$ gauge field in 2 + 1 dimensions with a total flux of ${\rm
Tr}\;\int F = 2 \pi k$ corresponds to a system of $k$ 0-branes bound
to $N$ 2-branes.  Generally, the energy of such a configuration is
minimized when the flux is distributed uniformly over the surface of
the 2-branes.  This is intuitively clear from the relative scaling of
the Yang-Mills energy $\int F^2$ and the 0-brane charge $F$.  For
example, if the size of the 0-brane were reduced by a factor of 2,
this would correspond to a doubling of the flux, and also to a
doubling of the integrated Yang-Mills energy.

A particularly simple example of a bound 0-brane and 2-brane is a
configuration on a torus $T^2$ with sides of unit length and flux $F =
2 \pi$.  This can be described explicitly in terms of a $U(1)$ bundle
over the torus with first Chern class $C_1= 1$ and connection
\[
A_2 (x_1,  x_2) = 2 \pi x_1.
\]
This choice of connection can be associated with overlap functions
describing the bundle which are trivial in the $x_2$ direction and
given by $\Omega_1 (x_2) = \exp (2 \pi ix_2)$ in the $x_1$ direction.
With this choice of connection, the bound state configuration of a
0-brane and a 2-brane can be related through T-duality in the $x_2$
direction to a 1-brane which is diagonally wound on the dual torus.
Because the boundary condition in the $x_2$ direction is trivial, we
can use an explicit description of T-duality through the relation
$X_2 = i \partial_2 + 2 \pi \alpha' A_2$
\cite{WT-compact,grt}.  The dual torus has dimensions $(1,4 \pi^2 \alpha')$,
and the configuration of the diagonal 1-brane is described by the
equation
\[
X_2 = 2 \pi \alpha' A_2= 4 \pi^2 \alpha' x_1.
\]
In an analogous fashion, a bound state of $k$ 0-branes and $N$
2-branes can be described as a configuration in $U(N)$ gauge theory on
the torus with constant flux $2 \pi k I/N$.

Although the bound 0+2 state can be described in Yang-Mills theory,
for a correct calculation of the energy of such a configuration it is
necessary to use the Born-Infeld action, which is described for an
abelian theory by (up to an overall constant)
\[
S = \sqrt{-\det (\eta_{\mu \nu} + 2 \pi \alpha' F_{\mu \nu})}.
\]
A complete definition of the nonabelian Born-Infeld action has not yet
been given.  For configurations in which all components of the field
strength commute, however, it is sufficient to take the above action
and trace over the gauge group.  (for recent discussions of the
problems when the fields do not commute, see \cite{Tseytlin,ht}).
The resulting energy when the fields commute is given by
\[
E = {\rm Tr}\;\sqrt{1 + 4 \pi^2 \alpha'^2 F^2}
\]
The term of order $\alpha'^2$ in an expansion of this expression gives
the Yang-Mills contribution to the energy.  For bound states of
$k$ 0-branes and $N$ 2-branes, the Born-Infeld energy is
\[
E = \sqrt{N^2 + (4 \pi^2 \alpha' k)^2}
\]
which is precisely the BPS bound on the energy of such a bound state.
This energy is less than the sum of the separate 0-brane and 2-brane
energies so that the state is truly bound \cite{Polchinski-TASI}.
This energy is of course also equal to the length of the dual
diagonally wrapped 1-brane.

\subsection{0-branes on 4-branes}
\label{sec:04}

Now let us consider a configuration of $k$ 0-branes bound to $N$
4-branes.  As discussed in \cite{Douglas,Vafa-instantons} such a
configuration corresponds to a $U(N)$ instanton with instanton number
\[
k =\frac{1}{8 \pi^2}\int {\rm Tr}\;  (F \wedge F) = C_2 -\frac{1}{2} C_1^2.
\]
Unlike the previous situation, a 0-brane on a 4-brane can be
contracted to a point without increasing or decreasing its energy.
This follows from the fact that the Yang-Mills energy $F^2$ and the
instanton number $F\wedge F$ both scale quadratically in $F$.  In
fact, the instanton can be shrunk to a point and then the 0-brane can
be removed from the 4-brane without changing the energy of the system.
This corresponds to the fact that the 0-brane and 4-brane are
marginally bound.

For a fixed instanton number $k$ on a compact manifold, the space of
gauge field configurations which minimize the Yang-Mills energy can be
associated with the moduli space of self-dual (or anti-self-dual,
depending upon the sign of $k$) fields $F =\pm *F$.  The usual way of
seeing this relation is to note that $ F^2 = F_+^2 + F_-^2$ where
$F_{\pm}$ are the self-dual and anti-self-dual parts of $F$, while $F
\wedge F = F_+^2 -F_-^2$.  There is another argument for this
conclusion, however, which generalizes more naturally to higher
dimensions, and which we now describe briefly.

If we are considering configurations on a compact manifold such as
$T^4$, subject to the constraint that $ \int F = 0$ on all 2-cycles,
while the instanton number is fixed to be $k$, then any gauge field
configuration which satisfies
\[
\frac{\delta}{ \delta F_{\mu \nu} (x)}  \left(
\int d^4  x \;
F_{\mu \nu}^2 -\lambda (
\frac{1}{8 \pi^2} \int d^4  x \; F \wedge F -k) \right),
\]
where $\lambda$ is a constant Lagrange multiplier, must satisfy the
Yang-Mills equations.  Thus, any solution of $F = \lambda *F$ with
$\lambda$ a constant must be a solution of the Yang-Mills equations.
Since $**F = F$, this relation can only hold for $\lambda = \pm
1$, giving the self-dual and anti-self-dual conditions respectively.

An explicit example of a 0 + 4 configuration with $N = k = 2$ on $T^4$
is given by a $U(2)$ bundle with $C_2 = 2$ with connection
\begin{eqnarray*}
A_1 & = &  0\\
A_2 & = &  2 \pi x_1\tau_3\\
A_3 & = &  0\\
A_4 & = &  2 \pi x_3 \tau_3
\end{eqnarray*}
where
\[
\tau_3 =\left(\begin{array}{cc}
1 & 0\\
0 & -1
\end{array} \right)
\]
is the usual Pauli matrix.  The field strength is given by
\[
F_{12} = F_{34} = 2 \pi \tau_3
\]
and is self-dual.

Describing $A$ as a connection on a
bundle with trivial boundary conditions in directions 2 and 4, we can
T-dualize in those directions, and we arrive at a system comprising a pair of
2-branes wrapped on $T^4$.  The embeddings of these 2-branes are given
by
\begin{eqnarray*}
X_2 (x_1, x_3) & = &  \pm 4 \pi^2 \alpha' x_1\\
X_4 (x_1, x_3) & = &  \pm 4 \pi^2 \alpha' x_3
\end{eqnarray*}
As discussed in \cite{bdl}, this is a configuration of 2-branes
intersecting at angles which preserves $1/4$ of the total
supersymmetry in the system.  One easy way to see this is that if we
scale the original dimensions $L_2,  L_4$ of the torus so that the dual
torus has all dimensions of equal length then the coefficients $4
\pi^2 \alpha'$ in the above embedding go to $1$.  In this case, the
2-branes are perpendicular and are in a $1/4$ supersymmetric
configuration.  

The preservation of supersymmetry in a system of 2-branes intersecting
at angles is equivalent to the (anti)-self-duality condition of the
dual gauge fields \cite{bdl}.  When this condition is satisfied, as in
the system just described, there are no tachyonic instabilities in the
string theory in the background of the intersecting 2-branes.  This is
equivalent to the fact that the corresponding gauge theory background
is stable and has no negative eigenvalues in the spectrum of
fluctuations.

To determine the energy of this system we can again use the
Born-Infeld action since the curvatures commute.  The energy of the
system is given by
\[
E = {\rm Tr}\; \sqrt{\det (\delta_{\mu \nu} + 2 \pi \alpha' F_{\mu
\nu})} = 2 \sqrt{(1 +  4 \pi^2 \alpha'^2 F_{12}^2)
(1 +  4 \pi^2 \alpha'^2 F_{34}^2)} = 2(1 + 16 \pi^4\alpha'^2)
\]
which is precisely the energy of two 4-branes and two 0-branes when
all the branes are separated to large distances.  Thus, this indeed
corresponds to a marginally bound state.  Note that the energy is also
equal to the sum of the areas of the two T-dual 2-branes.

\section{0-branes on 6-branes}

Now that we have reviewed the binding of 0-branes to 2-branes and
4-branes, let us consider the case of 0-branes and 6-branes.  As
pointed out in \cite{Polchinski-TASI}, it is not possible in general to
bind a 0-brane to a 6-brane without putting extra energy into the
system.  Both the short-range and long-range forces between 0-branes
and 6-branes are repulsive.  (For discussions of 0-brane/6-brane
scattering, see \cite{Lifschytz,Vijay-Finn2}).  In fact, imagine that we
could attach a 0-brane to $N$ infinite 6-branes by producing a  gauge
configuration with 
\[
\frac{1}{48 \pi^3}\int {\rm Tr}\; (F \wedge F \wedge F) = 1
\]
with vanishing $\int {\rm Tr}\; (F \wedge F)$ and $\int {\rm Tr}\; F$,
corresponding to an absence of 2-branes and 4-branes.  By scaling
\[
\tilde{F} (x) = \rho^{2} F (\rho x)
\]
we would get a new configuration with the same 0-brane charge but with
Yang-Mills energy $\tilde{E} = E/\rho^2$.  Thus, by taking $\rho
\rightarrow \infty$, the 0-brane can be shrunk to a point while
decreasing the total energy of the system.  This corresponds to the
fact that generically a 0-brane on a 6-brane will shrink to a point
and then will be pushed off the 6-brane.

In spite of these considerations, however, we now proceed to construct
an explicit gauge field configuration on $T^6$ which describes a system of
four 0-branes and four 6-branes.  This configuration satisfies the
classical equations of motion and is classically stable, at least to
quadratic order.  We begin by generalizing the argument given in
section \ref{sec:04} for solutions of the Yang-Mills equations with
fixed topology.  Imagine that we have a gauge field on a
bundle over $T^6$ with vanishing first and second Chern classes, but
with 0-brane charge $k$.  Any gauge field configuration which
satisfies for some fixed value of $\lambda$ the equation
\[
\frac{\delta}{ \delta F_{\mu \nu} (x)}  \left(
\int d^6  x \;
F_{\mu \nu}^2 -\lambda (
\frac{1}{48 \pi^3} 
\int d^6  x \; F \wedge F \wedge F -k) \right)
\]
must satisfy the Yang-Mills equations.  Thus, we can construct moduli
spaces of solutions to the Yang-Mills equations by finding solutions
to the equation
\begin{equation}
F = \lambda *(F \wedge F).
\label{eq:6-equation}
\end{equation}
We will now proceed to explicitly construct such a solution.

We choose the following
connection on $T^6$
\begin{eqnarray*}
A_1 =  0& \;\;\;\;\; &
A_2 =  2 \pi x_1 \mu_1\\
A_3 =  0& \;\;\;\;\; &
A_4 =  2 \pi x_3 \mu_2\\
A_5 =  0&\;\;\;\;\; &
A_6 =  2 \pi x_5 \mu_3
\end{eqnarray*}
where $\mu_i$ are the following generators of the $U (4)$ algebra
\[
\mu_1=\left(
\begin{array}{cccc}
 1 & 0 & 0 & 0\\
0 & 1 & 0 & 0\\
0 & 0 & -1 & 0\\
0 & 0 & 0 & -1\\
\end{array} \right)
\;\;\;\;\; 
\mu_2=\left(
\begin{array}{cccc}
 1 & 0 & 0 & 0\\
0 & -1 & 0 & 0\\
0 & 0 & -1 & 0\\
0 & 0 & 0 & 1\\
\end{array} \right)
\]
\[
\mu_3=\left(
\begin{array}{cccc}
 1 & 0 & 0 & 0\\
0 & -1 & 0 & 0\\
0 & 0 & 1 & 0\\
0 & 0 & 0 & -1\\
\end{array} \right)
\]
These generators have the properties that 
\begin{eqnarray}
{\rm Tr}\; \mu_i & = &  0\nonumber\\
{\rm Tr}\; \mu_i \mu_j & = &  0, \; \; {\rm for} \; i \neq j
\label{eq:relations}\\
\mu_i \mu_j & = & | \epsilon_{ijk} | \mu_k. \nonumber
\end{eqnarray}
The field strength associated with this connection is
\begin{equation}
F_{12}  =  2 \pi \mu_1\;\;\;\;\;
F_{34}  =  2 \pi \mu_2\;\;\;\;\;
F_{56}  = 2 \pi \mu_3
\label{eq:}
\end{equation}
{}From the above properties, we can easily see that
\[
\int d^4 x \; {\rm Tr}\; (F \wedge F) = 0
\]
around any 4-cycle on $T^6$ and that
\[
\int d^2x \; {\rm Tr}\;F = 0
\]
around any 2-cycle, while
\[
\int d^6x \; {\rm Tr}\; (F \wedge F \wedge F) = 4 \cdot 48 \pi^3
\]
corresponding to a charge of four 0-branes.
Furthermore, as a result of the last relation in (\ref{eq:relations}),
the field strength satisfies (\ref{eq:6-equation}) and therefore
satisfies the Yang-Mills equations.

By choosing boundary overlap functions defining the bundle which are
trivial in directions 2, 4, and 6, we can apply T-duality in those
three directions, and we get a resulting configuration of four 3-branes
wrapped on the dual $T^6$.  If we choose the length of dimensions 2, 4
and 6 in the original torus to be $L = 4 \pi^2 \alpha'$ then the dual
torus has unit length in all directions.  The four 3-branes wrapped on
this torus are described by the equations
\begin{eqnarray*}
X^{(1)}_2 (x_1, x_3, x_5)  = x_1 &
X^{(1)}_4 (x_1, x_3, x_5)  = x_3 &
X^{(1)}_6 (x_1, x_3, x_5)  = x_5 \\
X^{(2)}_2 (x_1, x_3, x_5)  = x_1 &
X^{(2)}_4 (x_1, x_3, x_5)  = -x_3 &
X^{(2)}_6 (x_1, x_3, x_5)  = -x_5 \\
X^{(3)}_2 (x_1, x_3, x_5)  = -x_1 &
X^{(3)}_4 (x_1, x_3, x_5)  = -x_3 &
X^{(3)}_6 (x_1, x_3, x_5)  = x_5 \\
X^{(4)}_2 (x_1, x_3, x_5)  = -x_1 &
X^{(4)}_4 (x_1, x_3, x_5)  = x_3 &
X^{(4)}_6 (x_1, x_3, x_5)  = -x_5 
\end{eqnarray*}
These 3-branes are all wrapped diagonally and each has volume $2^{3/2}$.
The 3-branes are in a configuration which breaks all supersymmetries.
This can be verified directly; however, a simple way to see this is to
observe that under a coordinate transformation
\begin{eqnarray*}
\tilde{x}_1 = x_1 + x_2 & 
\tilde{x}_3 = x_3 + x_4 & 
\tilde{x}_5 = x_5 + x_6 \\
\tilde{x}_2 = x_1 - x_2 & 
\tilde{x}_4 = x_3 - x_4 & 
\tilde{x}_6 = x_5 - x_6 
\end{eqnarray*}
the 3-branes are wrapped parallel to the sets of coordinates
\[
(\tilde{1}\tilde{3}\tilde{5}), (\tilde{1}\tilde{4}\tilde{6}), 
(\tilde{2}\tilde{4}\tilde{5}), (\tilde{2}\tilde{3}\tilde{6})
\]
This type of configuration of intersecting 3-branes was previously
considered in \cite{Vijay-Finn} where a similar configuration was used
to study black holes in 4D.  As discussed in that paper, for a system
of intersecting 3-branes of this type, for half of the 16 choices of
orientations for the 3-branes the configuration is 1/8 supersymmetric
and for the other half all supersymmetries are broken.  The
configuration we have constructed here corresponds to a choice of
orientations which breaks all supersymmetries.
Let us briefly review the argument for
the breaking of supersymmetry in this situation.  A priori, each brane should
break 1/2 of the supersymmetry.  However, the conditions on an
unbroken supersymmetry given by the four branes are
\begin{eqnarray*}
\tilde{\epsilon}  =  M_1 \epsilon =\tilde{\Gamma}_{0} \tilde{\Gamma}_{1}
\tilde{\Gamma}_{3} \tilde{\Gamma}_{5} \epsilon
& = &M_2 \epsilon =\tilde{\Gamma}_{0} \tilde{\Gamma}_{1}
\tilde{\Gamma}_{4} \tilde{\Gamma}_{6} \epsilon \\
   =M_3 \epsilon =  \tilde{\Gamma}_{0} \tilde{\Gamma}_{2}
\tilde{\Gamma}_{4} \tilde{\Gamma}_{5} \epsilon
& = &M_4 \epsilon =\tilde{\Gamma}_{0} \tilde{\Gamma}_{2}
\tilde{\Gamma}_{3} \tilde{\Gamma}_{6} \epsilon 
\end{eqnarray*}
The matrices $M_i$ satisfy the relation
\[
(M_1)^{-1} M_2 = -(M_3)^{-1} M_4,
\]
which guarantees that all supersymmetries must be broken.  One might
hope that by simply changing the orientation of one of the branes one
could reach a supersymmetric configuration; however, in addition to
introducing 2-brane and 4-branes charges into the system, this would invert
the orientation of one of the 6-branes in the original configuration,
making it impossible to describe this system in terms of a gauge
theory on the 6-brane world-volume.

Since all supersymmetries are broken in this configuration we expect
that the energy should exceed the minimal BPS energy for a 0 + 6
system.  Indeed, to determine the energy of this configuration we can
again use the Born-Infeld formula appropriate for diagonal field
strengths.  The total energy of the configuration is
\[
E = (4 \pi^2 \alpha')^{3} 4 \sqrt{(1 + 1)^3} = 
8 \sqrt{2} (4 \pi^2 \alpha')^{3}
\]
since the fluxes are all equal to $(2 \pi \alpha')^{-1}$ while the
volume of $T^6$ is $ (4 \pi^2 \alpha')^{3}$ with the sides of the
torus as above.  On the other hand, the energy of four 0-branes and
four 6-branes if all are separate is given by
\[
E_{\rm sep} = 8  (4 \pi^2 \alpha')^{3}
\]
(where, again, we have normalized 6-branes to have an energy equal to
their volume).  Since the separated energy is lower than the energy in
the combined configuration, this state is not a bound state in the
standard terminology.  However, the 0-branes have nonetheless adhered
to the 6-branes in a relatively stable fashion.

We can also consider the energy in the dual picture.  In this case the
total volume contained in the diagonally wrapped branes is $E = 8
\sqrt{2}$ since there are four branes each with volume $2 \sqrt{2}$.  The
brane configuration can be characterized by the 3-brane charges along
all homology 3-cycles.  The charges in all directions other than the
$135$ and $246$ cycles cancel, so that the configuration has the same
charges as a system of four branes wrapped around the $135$ cycle and
four branes wrapped around the $246$ cycle.  Such branes would have
unit volume and so the total volume of such a perpendicular
configuration would be 8, less than the volume of the diagonally
wrapped branes.  

The configuration we have constructed here is stable, at least to
quadratic order, with respect to fluctuations in the gauge field.
{}From the string point of view in the dual 3-brane picture this follows
from the fact that the strings stretching between each pair of branes
are like strings stretching between two perpendicular 2-branes, since
each pair of 3-branes has a single direction in which they are
coincident.  Thus, there are no tachyonic instabilities in the string
theory spectrum around this background.  Just as in the 4D case, such
an instability would correspond to an unstable direction in the
original gauge theory description.  Although there are no unstable
directions at quadratic order, there are flat directions corresponding
to parameters in the moduli space of solutions to $F = \lambda*(F
\wedge F)$.  In the dual 3-brane picture these flat directions
correspond to the massless fields in the string spectrum arising from
strings stretching between pairs of 3-branes.

\section{0-branes on 8-branes}

We  now briefly describe an analogous construction for a system of
eight 0-branes and eight 8-branes on $T^8$.  Just as for the
6-brane, a 0-brane attached to an 8-brane will generically contract to
a point.  This follows from the relative scaling
of $F^2$ and $F \wedge F \wedge F \wedge F$.  We would expect  a
solution of the classical Yang-Mills equations corresponding to an
adhered state to satisfy the relation $F = \lambda *(F \wedge F \wedge
F)$.  Indeed, considering a system of eight 0-branes and  eight
8-branes, such a configuration can be constructed in terms of a
constant curvature connection.  The four independent components of F
can be taken to be proportional to the matrices
\begin{eqnarray*}
F_{12} & = &  \frac{1}{2 \pi \alpha'} 
 \;{\rm Diag} (1, 1, 1, 1, -1, -1, -1, -1)\\
F_{34} & = &  \frac{1}{2 \pi \alpha'}  
\;{\rm Diag} (1, 1, -1, -1, 1, 1, -1, -1)\\
F_{56} & = &  \frac{1}{2 \pi \alpha'}  
\;{\rm Diag} (1, -1, 1, -1, 1, -1, 1, -1)\\
F_{78} & = &  \frac{1}{2 \pi \alpha'}  
\;{\rm Diag} (1, -1, -1, 1, -1, 1, 1, -1)
\end{eqnarray*}

Scaling the dimensions of directions $2, 4, 6$ and 8 so that after
T-duality in those directions we get a $T^8$ with all sides of unit
length, we see that the dual configuration to this field strength
corresponds to a system of eight 4-branes wrapped diagonally on
$T^8$.  These 4-branes are arranged so that every pair is
perpendicular in either four or eight dimensions.  This configuration
again breaks all the supersymmetries of the theory.

The energy of this configuration is precisely twice the energy that
the 0-branes and 8-branes would have if they were completely
separated.  This can be seen easily in the dual picture, where we have
eight 4-branes with volume  $4 = (\sqrt{2})^4$.  These 4-branes are
equivalent in homology to a system of eight 4-branes wrapped in
dimensions $1357$ and eight 4-branes wrapped in dimensions $2468$,
which would each have a unit volume.  However, as for the 0 + 6
system, this configuration is stable classically, at least to
quadratic order.

A system of separated 0-branes and 8-branes can preserve some
supersymmetry \cite{Polchinski-TASI}, unlike the situation with
0-branes and 6-branes.  In fact, it is believed that 0-branes and
8-branes can form marginally bound states.  Nonetheless, in view of
the fact that 0-branes tend to become pointlike when embedded in
infinite 8-branes, at least at the level of Yang-Mills or Born-Infeld
theory, it would be interesting to study configurations of these
objects further.  Recently eight dimensional Yang-Mills theory has
been discussed in a related context
\cite{bks}.  A discussion of 0-branes on 8-branes was also given in
\cite{Horava}, and bound states of 0-branes and 8-branes in type I'
theory were described in \cite{Kachru-Silverstein}.  In order to
achieve a full understanding of bound states of 0-branes and 8-branes
in type II string theory it will probably also be necessary to
incorporate some aspect of the phenomenon in which a 0-brane passing
through an 8-brane produces an extra string
\cite{Hanany-Witten,bdg,dfk}.

\section{Conclusions}

We have constructed a configuration of four 0-branes and four 6-branes
which have adhered together to form a  state with no classical
instability at quadratic order.  This
configuration is dual to a system of four 3-branes wrapped diagonally
on $T^6$.  The state breaks all supersymmetry.  We also constructed an
analogous configuration of eight 0-branes and eight 8-branes on $T^8$
which is dual to a system of eight intersecting 4-branes.

One application of the construction described here is to dualize these
configurations to get constructions of the 6-brane and 8-brane in
M(atrix) theory along the lines of known constructions of 2-branes and
4-branes \cite{BFSS,grt,bss}.  Suggestions related to the construction
of 6-branes or 8-branes in M(atrix) theory were also discussed
recently in \cite{bss,bc,Horava}.  There are a number of tricky issues
related to the interpretation of these objects; nonetheless, having an
explicit construction of such a configuration might help to understand
some of the puzzling aspects of these systems.

Since the configurations which have been described here break
supersymmetry and have extra binding energy beyond the energy of their
constituents, they presumably do not correspond to truly stable brane
configurations.  However, because they are stable classically at
quadratic order, they may correspond to some kind of long-lived
resonances composed of 0-branes and 6-branes or 8-branes.  Such
metastable states are expected from the supergravity point of view in
the 0 + 6 case when the charges are large \cite{Sheinblatt}, but the
existence of the configurations constructed here indicates that
perhaps a long-lived configuration of 0-branes and 6-branes can be
found even for small numbers of branes.  It would be interesting to
study further the effects of quantum corrections on these
configurations.  It would also be interesting to understand better the
classical moduli spaces of Yang-Mills solutions in which these
configurations lie.  Presumably these correspond to six-dimensional
and eight-dimensional analogues of the well-studied moduli space of
self-dual connections on four-manifolds.  Because the relevant
equations in these cases are nonlinear, this is probably a more
difficult mathematical problem.  However, at least it is
straightforward to compute the dimensions of these moduli spaces;
given a point in the moduli space corresponding to a constant
curvature connection, the dimension of the moduli space can be
computed by finding the number of physical zero-modes around the given
gauge theory background.  Generalizing to a space of solutions with
arbitrary numbers of 0-branes and 6-branes, for example, the
appropriate moduli space of Yang-Mills solutions should correspond to
a space whose dimensionality is related to the entropy of the type of
0 + 6 black hole configuration considered in
\cite{Khuri-Ortin,Sheinblatt}.

\section*{Acknowledgements}
I would like to thank V.\ Balasubramanian, O.\ Ganor, A.\ Hashimoto,
S. Ramgoolam, I.\ M.\ Singer, K.\ Swanson, L.\ Thorlacius and E.\
Weinstein for helpful conversations.  This work was supported by the
National Science Foundation (NSF) under contract PHY96-00258.


\end{document}